\documentclass[aps,prd,twocolumn,superscriptaddress,showpacs]{revtex4}
\usepackage[dvips]{graphicx}
\usepackage{amsmath,amssymb,times}

\newcommand{\bea}{\begin{eqnarray}}
\newcommand{\eea}{\end{eqnarray}}

\newcommand{\la}{\left\langle}

\def\la{\mathrel{\mathpalette\fun <}}
\def\ga{\mathrel{\mathpalette\fun >}}
\def\fun#1#2{\lower3.6pt\vbox{\baselineskip0pt\lineskip.9pt
  \ialign{$\mathsurround=0pt#1\hfil##\hfil$\crcr#2\crcr\sim\crcr}}}

\DeclareSymbolFont{boldletters}{OML}{cmm} {b}{it}
\DeclareSymbolFontAlphabet{\mathbit}{boldletters}
\DeclareMathSymbol{\alpha}{\mathalpha}{letters}{"0B}
\DeclareMathSymbol{\beta}{\mathalpha}{letters}{"0C}
\DeclareMathSymbol{\gamma}{\mathalpha}{letters}{"0D}
\DeclareMathSymbol{\delta}{\mathalpha}{letters}{"0E}
\DeclareMathSymbol{\epsilon}{\mathalpha}{letters}{"0F}
\DeclareMathSymbol{\zeta}{\mathalpha}{letters}{"10}
\DeclareMathSymbol{\eta}{\mathalpha}{letters}{"11}
\DeclareMathSymbol{\theta}{\mathalpha}{letters}{"12}
\DeclareMathSymbol{\iota}{\mathalpha}{letters}{"13}
\DeclareMathSymbol{\kappa}{\mathalpha}{letters}{"14}
\DeclareMathSymbol{\lambda}{\mathalpha}{letters}{"15}
\DeclareMathSymbol{\mu}{\mathalpha}{letters}{"16}
\DeclareMathSymbol{\nu}{\mathalpha}{letters}{"17}
\DeclareMathSymbol{\xi}{\mathalpha}{letters}{"18}
\DeclareMathSymbol{\pi}{\mathalpha}{letters}{"19}
\DeclareMathSymbol{\rho}{\mathalpha}{letters}{"1A}
\DeclareMathSymbol{\sigma}{\mathalpha}{letters}{"1B}
\DeclareMathSymbol{\tau}{\mathalpha}{letters}{"1C}
\DeclareMathSymbol{\upsilon}{\mathalpha}{letters}{"1D}
\DeclareMathSymbol{\phi}{\mathalpha}{letters}{"1E}
\DeclareMathSymbol{\chi}{\mathalpha}{letters}{"1F}
\DeclareMathSymbol{\psi}{\mathalpha}{letters}{"20}
\DeclareMathSymbol{\omega}{\mathalpha}{letters}{"21}
\DeclareMathSymbol{\varepsilon}{\mathalpha}{letters}{"22}
\DeclareMathSymbol{\vartheta}{\mathalpha}{letters}{"23}
\DeclareMathSymbol{\varpi}{\mathalpha}{letters}{"24}
\DeclareMathSymbol{\varrho}{\mathalpha}{letters}{"25}
\DeclareMathSymbol{\varsigma}{\mathalpha}{letters}{"26}
\DeclareMathSymbol{\varphi}{\mathalpha}{letters}{"27}
\DeclareMathSymbol{\Gamma}{\mathalpha}{letters}{"00}
\DeclareMathSymbol{\Delta}{\mathalpha}{letters}{"01}
\DeclareMathSymbol{\Theta}{\mathalpha}{letters}{"02}
\DeclareMathSymbol{\Lambda}{\mathalpha}{letters}{"03}
\DeclareMathSymbol{\Xi}{\mathalpha}{letters}{"04}
\DeclareMathSymbol{\Pi}{\mathalpha}{letters}{"05}
\DeclareMathSymbol{\Sigma}{\mathalpha}{letters}{"06}
\DeclareMathSymbol{\Upsilon}{\mathalpha}{letters}{"07}
\DeclareMathSymbol{\Phi}{\mathalpha}{letters}{"08}
\DeclareMathSymbol{\Psi}{\mathalpha}{letters}{"09}
\DeclareMathSymbol{\Omega}{\mathalpha}{letters}{"0A}

\begin{document}
\preprint{SAGA-HE-255-09}
\title{Average phase factor in the PNJL model}

\author{Yuji Sakai}
\email[]{sakai@phys.kyushu-u.ac.jp}
\affiliation{Department of Physics, Graduate School of Sciences, Kyushu University,
             Fukuoka 812-8581, Japan}

\author{Takahiro Sasaki}
\email[]{sasaki@phys.kyushu-u.ac.jp}
\affiliation{Department of Physics, Graduate School of Sciences, Kyushu University,
             Fukuoka 812-8581, Japan}

\author{Hiroaki Kouno}
\email[]{kounoh@cc.saga-u.ac.jp}
\affiliation{Department of Physics, Saga University,
             Saga 840-8502, Japan}

\author{Masanobu Yahiro}
\email[]{yahiro@phys.kyushu-u.ac.jp}
\affiliation{Department of Physics, Graduate School of Sciences, Kyushu University,
             Fukuoka 812-8581, Japan}

\date{\today}

\begin{abstract}
The average phase factor $\langle e^{2i\theta}\rangle$ 
of the QCD determinant is evaluated 
at finite quark chemical potential ($\mu_{\rm q}$)
with the two-flavor version of the 
Polyakov-loop extended Nambu-Jona-Lasinio (PNJL) model 
with the scalar-type eight-quark interaction. 
For $\mu_{\rm q}$ larger than half the pion mass $m_{\pi}$ at vacuum, 
$\langle e^{2i\theta}\rangle$ is finite 
only when the Polyakov loop is larger than $\sim 0.5$, 
indicating that lattice QCD is feasible only in the deconfinement 
phase. 
A critical endpoint (CEP) lies in the region 
of $\langle e^{2i\theta}\rangle =0$. 
The scalar-type eight-quark interaction makes it shorter 
a relative distance of the CEP to the boundary of the region. 
For $\mu_{\rm q} < m_{\pi}/2$, 
the PNJL model with dynamical mesonic fluctuations can reproduce 
lattice QCD data below the critical temperature. 
\end{abstract}

\maketitle

\section{Introduction}
\label{Introduction}
The thermodynamics of quantum chromodynamics (QCD) is well 
defined, since QCD is renormalizable and parameter free. 
Nevertheless, the thermodynamics 
is not understood 
at lower temperature ($T$) 
because of its nonperturbative nature. 
The thermodynamics is closely related to not only natural 
phenomena such as compact stars and the early universe but also 
laboratory experiments such as relativistic heavy-ion collisions. 
Lattice QCD (LQCD) is a first-principle calculation, 
but it has the well-known sign problem 
when the quark-number chemical potential ($\mu_{\rm q}$) 
is real; for example, 
see Ref.~\cite{Kogut}. 
So far, several approaches have been proposed to circumvent the difficulty; 
for example, the reweighting method~\cite{Fodor}, 
the Taylor expansion method~\cite{Allton} and 
the analytic continuation from imaginary $\mu_{\rm q}$ 
to real $\mu_{\rm q}$~\cite{FP,Elia,Elia2,Chen}. 
However, these are still far from perfection particularly at 
$\mu_{\rm q}/T \ga 1$.

The success of the approaches is linked to how difficult 
the sign problem is.  
As a good index of the difficulty, one can consider 
the average of the phase factor 
\begin{eqnarray}
e^{2i\theta}
=\frac{\det(D+\mu_{\rm q}\gamma_0+m)}{\det(D+\mu_{\rm q}\gamma_0+m)^*} 
\end{eqnarray}
of the Fermion determinant. 
If the average of the phase factor is much smaller than 1, this 
means that there are severe cancellations 
in the path integral of the QCD partition function. 
In this situation, LQCD simulations are not feasible.

The average is obtained by taking 
the expectation value of the phase factor in the 
phase-quenched theory in which the Fermion determinant is replaced 
by the absolute value. 
In the two-flavor case, the average is 
\begin{eqnarray}
\langle e^{2i\theta}\rangle=\frac{Z_{1+1}}{Z_{1+1^*}}, 
\label{eq2}
\end{eqnarray}
where $Z_{1+1}$ stands for 
the partition function of the ordinary two-flavor theory 
and $Z_{1+1^*}$ represents that of the two-flavor phase-quenched theory 
in which one of two flavors is changed into a conjugate flavor. 
For comparison of the 1+1$^*$ system with the 1+1 system, 
let us introduce the modified isospin 
chemical potential $\mu_{\rm I}$ related to 
the isospin chemical potential $\mu_{\rm iso}$ 
as $\mu_{\rm I}=\mu_{\rm iso}/2$. 
When the 1+1 system has a value of $\mu_{q}$, 
the 1+1$^*$ system possesses the same value of $\mu_{\rm I}$.

It is not easy to calculate the average phase factor with LQCD even for 
small $\mu_{\rm q}/T$. 
Actually, several LQCD results on the average phase factor 
are spotted; see Ref.~\cite{Danzer} and references therein. 
It is then important to make a systematic analysis on the phase factor 
by using effective theories. This was done by 
the chiral perturbation theory~\cite{Splittorff,Bloch}. 
The result is consistent with LQCD one~\cite{Danzer} when $T$ 
is lower than the critical one $T_{\rm c}$. However, 
the theory is not valid for $T > T_{\rm c}$.

Recently, the average phase factor at $T > T_{\rm c}$ was calculated by 
the random matrix theory~\cite{Han} and 
the Nambu--Jona-Lasinio (NJL) model~\cite{Anderson}. 
When the saddle-point approximation is applied to 
the path integrals in the partition functions 
$Z_{1+1}$ and $Z_{1+1^*}$, the average phase factor can be described by 
\begin{eqnarray}
\label{MFgeneral}
\langle e^{2i\theta}\rangle \approx 
\frac{\sqrt{\det H_{1+1^*}}}{\sqrt{\det H_{1+1}}}
e^{-\beta V (\Omega_{1+1}-\Omega_{1+1^*})},
\end{eqnarray}
where $\beta=1/T$, $\Omega$ is 
the thermodynamic potential at mean field level and 
$H$ is the Hessian matrix showing static fluctuations (SF) 
at the saddle point. 
The average phase factor thus obtained 
is dominated not by the exponential factor 
$e^{-\beta V (\Omega_{1+1}-\Omega_{1+1^*})}$ but by 
the SF factor 
${\sqrt{\det H_{1+1^*}}}/{\sqrt{\det H_{1+1}}}$~\cite{Anderson}, 
because $\Omega_{1+1}=\Omega_{1+1^*}$ in the normal phase with no pion 
condensate and the SF factor is zero 
in the pion condensate phase; 
this will be explained explicitly in subsection \ref{Average phase factor}. 
Thus, the average phase factor should be calculated 
with the mean field (MF) approximation  plus SF corrections. 
This framework is referred to as MF+SF in the present paper.

The NJL model describes the chiral phase transition
~\cite{NJ1,AY,Kle,HK1,Sakaguchi,Kashiwa1}, 
but not the deconfinement transition. 
The Polyakov-loop extended Nambu--Jona-Lasinio (PNJL) 
model~\cite{Meisinger,Fukushima,Fukushima2,Ghos,Megias,Ratti,
Rossner,Ciminale,Schaefer,Zhang,Mukherjee,Costa,Kashiwa2,Fu,Hell,
Sakai1,Sakai2,Sakai3,Xiong,Sasaki-T} was constructed to treat 
both the transitions simultaneously. 
Very recently, the PNJL model was shown to be successful in 
reproducing LQCD data for imaginary quark chemical 
potential~\cite{Sakai1,Sakai2}, imaginary quark and isospin 
chemical potentials~\cite{Sakai3} and real 
isospin chemical potential~\cite{Sasaki-T}. 
Thus, the PNJL model is one of the most reliable models.

In this paper, we evaluate the average phase factor 
by the PNJL model in the MF+SF framework and investigate 
a relation between the Polyakov-loop and the average phase factor. 
For thermal systems at $\mu_{\rm I}=\mu_{\rm q}=0$ and at 
$\mu_{\rm I}>0$ and $\mu_{\rm q}=0$, 
the scalar-type eight-quark interaction is essential for the PNJL model 
to reproduce LQCD data~\cite{Sakai2,Sasaki-T}. 
We then analyze an effect of the eight-quark interaction on the 
average phase factor, too. 
Finally, we consider dynamical mesonic fluctuations (DF), 
instead of static mesonic fluctuations, 
to compare the PNJL results with LQCD data.

This paper is organized as follows. 
In Sec. II, we recapitulate the PNJL model and a way of 
calculating the average phase factor.  
The numerical results are shown in Sec. III. 
Effects of the Polyakov loop, the eight-quark interaction, 
static fluctuations and dynamical mesonic fluctuations are investigated. 
Section IV is devoted to summary.

\section{Formalism}
\label{Formalism} 
In this section, we first explain the PNJL model in the MF level 
for the case of finite $\mu_{\rm q}$ and $\mu_{\rm I}$ and 
treat SF in order to 
evaluate the average phase factor. 

\subsection{PNJL model}
\label{PNJL}
The Lagrangian of the two-flavor PNJL model in Euclidean spacetime is 
\begin{align}
\mathcal{L}&=&
\bar{q}(\gamma_\nu D^\nu - \gamma_4{\hat \mu} + {\hat m}_0 )q 
+G_{\rm s}\left[
(\bar{q}q)^{2}+(\bar{q}i\gamma_{5} \vec{\tau} q)^{2}
\right]
 \notag \\
&+&
G_{\rm s8}\left[
(\bar{q}q)^{2}+(\bar{q}i\gamma_{5} \vec{\tau} q)^{2}
\right] ^{2}
- {\cal U}(\Phi [A],{\Phi} [A]^*,T), 
\label{eq:E1}
\end{align}
where $D^\nu=\partial^\nu+iA^\nu$ and 
$A^\nu=\delta^{\nu}_{0}gA^0_a{\lambda^a\over{2}}$
with the gauge field $A^\nu_a$, 
the Gell-Mann matrix $\lambda_a$ and the gauge coupling $g$. 
The coupling constant $G_{\rm s}$ ($G_{\rm s8}$) represents a strength of 
the scalar-type 
four-quark (eight-quark) interaction. 
The Polyakov potential ${\cal U}$ of (\ref{eq:E13}) 
is a function of the Polyakov loop $\Phi$ and its Hermitian 
conjugate $\Phi^*$.

The chemical potential matrix ${\hat \mu}$ is 
defined by ${\hat \mu}={\rm diag}(\mu_u, \mu_d)$ with 
the $u$-quark ($d$-quark) number chemical potential $\mu_{u}$ ($\mu_{d}$), 
while ${\hat m}_0={\rm diag}(m_0, m_0)$. 
The chemical potential matrix is described by 
$\mu_{\rm q}$ and $\mu_{\rm I}$ as 
\begin{align}
{\hat \mu}=\mu_{\rm q} \tau_0 + \mu_{\rm I} \tau_3, 
\end{align}
where $\tau_0$ is the unit matrix and $\tau_i$ ($i=1, 2, 3$) 
are the Pauli matrices in flavor space. 
The $\mu_{\rm q}$ and $\mu_{\rm I}$ are related to 
the baryon and isospin chemical potentials, 
$\mu_{\rm B}$ and $\mu_{\rm iso}$, 
coupled respectively to the baryon charge ${\bar B}$ and to the isospin 
charge ${\bar I_3}$ as 
\begin{align}
\mu_{\rm q}=\frac{\mu_{u}+\mu_{d}}{2}=\frac{\mu_{\rm B}}{3},
~~\mu_{\rm I}=\frac{\mu_{u}-\mu_{d}}{2}=\frac{\mu_{\rm iso}}{2} . 
\end{align}
For later convenience, we use $\mu_{\rm I}$ instead of 
$\mu_{\rm iso}$ and call it the isospin chemical potential simply. 
The $1$+$1$ and the $1$+$1^*$ theory correspond to taking 
$(\mu_{\rm q}, \mu_{\rm I})=(\mu_{\rm q}, 0)$ and 
$(0, \mu_{\rm q})$, respectively, in $\mathcal{L}$ of \eqref{eq:E1}.  
In the limit of $m_0=\mu_{\rm I}=0$, 
the PNJL Lagrangian has the $SU_{\rm L}(2) \times SU_{\rm R}(2)
\times U_{\rm v}(1) \times SU_{\rm c}(3)$  symmetry. 
For $m_0 \neq 0$ and $\mu_{\rm I} \neq 0$, 
it is reduced to $U_{\rm I_3}(1) \times U_{\rm v}(1) \times SU_{\rm c}(3)$.

The Polyakov-loop operator $\hat{\Phi}$ and its Hermitian conjugate 
$\hat{\Phi}^{\dagger}$ are defined as
\begin{eqnarray}
\hat{\Phi}        &=& {1\over 3} {\rm Tr} L ,~~~~
\hat{\Phi}^{*}  = {1\over 3} {\rm Tr}L^\dag ,
\end{eqnarray}
with
\begin{eqnarray}
L({\bf x})  &=& {\cal P} \exp\Bigl[
                {i\int^\beta_0 d \tau A_4({\bf x},\tau)}\Bigr],
\end{eqnarray}
where ${\cal P}$ is the path ordering and $A_4 = i A^0 $. 
In the Polyakov gauge, $L$ can be written in a diagonal form 
in color space~\cite{Fukushima}: 
\begin{align}
L =  e^{i \beta (\phi_3 \lambda_3 + \phi_8 \lambda_8)}. 
\label{eq:E6}
\end{align}
In the MF level, $\phi_3$ and $\phi_8$ 
are treated as classical variables~\cite{Rossner}.

The spontaneous breakings of the chiral and the $U_{\rm I_3}(1)$ symmetry 
are described, respectively, 
by the chiral condensate $\sigma = \langle \bar{q}q \rangle$ and the charged 
pion condensate~\cite{Zhang}
\begin{align}
\pi^{\pm}=\frac{\pi}{\sqrt{2}}e^{\pm i \alpha}
=\langle \bar{q}i \gamma_5 \tau_{\pm}q \rangle.
\label{charged-pion}
\end{align}
Since the phase $\alpha$ represents the direction 
of the $U_{\rm I_3}(1)$ symmetry breaking, 
we take $\alpha=0$ for convenience. The pion 
condensate is then expressed by 
\begin{align}
\pi=\langle \bar{q}i \gamma_5 \tau_{1}q \rangle.
\label{pion}
\end{align}
Making the MF approximation~\cite{Kashiwa1,Zhang}, 
one can obtain the MF Lagrangian as 
\begin{align}
 {\cal L}_{\rm MF}  &=& {\bar q}(\gamma_\nu D^\nu - \gamma_4{\hat \mu} + 
              M\tau_0 + N i \gamma_5 \tau_{1})q  ~~~~~~\notag\\
            &\hspace{3mm}&  + G_{\rm s}[\sigma^2 +\pi^2] 
              + 3G_{\rm s8}(\sigma ^{2} + \pi^{2})^{2}
              + {\cal U} \quad
             \label{MF-L}
\end{align}
with 
\begin{eqnarray}
M&=&m_{0} - 2[G_{\rm s}+2G_{\rm s8}(\sigma ^{2}+\pi ^{2})] \sigma,~~\notag\\
N&=&- 2[G_{\rm s}+2G_{\rm s8}(\sigma ^{2}+\pi ^{2})] \pi. 
\end{eqnarray}
The quark propagator ${\cal S}$ in the MF level, that has 
off-diagonal elements in flavor space, is obtained by 
\begin{eqnarray}
{\cal S}^{-1}(p)
=\left(\begin{array}{cc}
\gamma_{\nu}P^{\nu}-\mu_u\gamma_4+M&-i\gamma_5N\\
i\gamma_5N&\gamma_{\nu}P^{\nu}-\mu_d\gamma_4+M
\end{array}\right),
\label{quarkpro}
\end{eqnarray}
where $P^{\nu}=p^{\nu}-A^{\nu}$. 
Performing the path integral in the PNJL partition function 
\begin{align}
Z_{\rm MF}=\int Dq D\bar{q}
\exp\left[ - \int d^4 x {\cal L}_{\rm MF} \right] , 
\label{PNJL-Z}
\end{align}
we can get the thermodynamic potential  
(per unit volume), 
\begin{align}
\Omega_{\rm MF} &=-T\ln(Z_{\rm MF})/V
= -2\sum_{i=\pm}\int \frac{d^3{\rm p}}{(2\pi)^3}
         \Bigl[ 3 E_{i}({\rm p}) \nonumber\\
       & + \frac{1}{\beta}
         \ln~ [1 + 3(\Phi+\Phi^{*} e^{-\beta E_{i}^-({\bf p})}) 
        e^{-\beta E_{i}^-({\bf p})}+ e^{-3\beta E_{i}^- ({\bf p})}]
         \nonumber\\
       & + \frac{1}{\beta} 
           \ln~ [1 + 3(\Phi^{*}+{\Phi e^{-\beta E_{i}^+({\bf p})}}) 
            e^{-\beta E_{i}^+({\bf p})}+ e^{-3\beta E_{i}^+({\bf p})}]
           \Bigl]\nonumber\\
       & +G_{\rm s}[\sigma^2 +\pi^2]
       + 3G_{\rm s8}(\sigma ^{2} + \pi^{2})^{2}
       +{\cal U} 
\label{eq:E12-pi} 
\end{align}
with 
\bea
E_{\pm}^\pm({\rm p})=E_{\pm}({\rm p})\pm \mu_{\rm q}, 
\eea
where $E_{\pm}({\rm p})=\sqrt{(E({\rm p})\pm\mu_{\rm I})^2+N^2}$ 
and $E({\rm p})=\sqrt{{\bf p}^2+M^2}$. 
The mometum integral in (\ref{eq:E12-pi}) is regularized by 
the three-dimensional cutoff $\Lambda$.

We use ${\cal U}$ of Ref.~\cite{Rossner}:
\begin{align}
&{\cal U} = T^4 \Bigl[-\frac{a(T)}{2} {\Phi}^*\Phi\notag\\
      &~~~~~+ b(T)\ln(1 - 6{\Phi\Phi^*}  + 4(\Phi^3+{\Phi^*}^3)
            - 3(\Phi\Phi^*)^2 )\Bigr], \label{eq:E13}\\
&a(T)   = a_0 + a_1\Bigl(\frac{T_0}{T}\Bigr)
                 + a_2\Bigl(\frac{T_0}{T}\Bigr)^2,
 ~~~b(T)=b_3\Bigl(\frac{T_0}{T}\Bigr)^3 .  
 \label{eq:E14}
\end{align}
The parameters of ${\cal U}$ are adjusted to LQCD data 
in the heavy-quark (pure-gauge) limit~\cite{Boyd,Kaczmarek}; 
the resultant parameter set is shown in Table I.  
In the limit, the Polyakov potential yields 
a first-order deconfinement phase transition at 
$T=T_0$. Since the first-order transition takes place at 
$T=270$~MeV in LQCD, $T_0$ is often set to 270~MeV in the PNJL 
calculation. 
In the light-quark case, however, 
the PNJL calculation with this value of $T_0$ yields a 
larger value of $T_\mathrm{c}$ than 
the full LQCD prediction 
$T_\mathrm{c}=173$~MeV~\cite{Karsch3,Karsch4,Kaczmarek2}. 
Therefore, we rescale $T_0$ to 200~MeV
to reproduce $T_\mathrm{c}=173$~MeV~\cite{Sakai2}.

\begin{table}[h]
\begin{center}
\begin{tabular}{llllll}
\hline
~~~~~$a_0$~~~~~&~~~~~$a_1$~~~~~&~~~~~$a_2$~~~~~&~~~~~$b_3$~~~~~
\\
\hline
~~~~3.51 &~~~~-2.47 &~~~~15.2 &~~~~-1.75\\
\hline
\end{tabular}
\caption{
Summary of the parameter set in the Polyakov-potential sector 
determined in Ref.~\cite{Rossner}. 
All parameters are dimensionless. 
}
\end{center}
\end{table}

In the NJL sector, two parameter sets are taken; in the first set 
$G_{\rm s8}$ is finite, while in the second set it is zero. 
The first set is 
$\Lambda =0.6315$~GeV, $G_{\rm s}=4.673$~[GeV$^{-2}]$, $G_{\rm s8}=452.12$~[GeV$^{-8}]$ and $m_0=5.5$~MeV. 
This set reproduces not only the pion decay constant $f_{\pi}=93.3$~MeV  
and the pion mass $m_{\pi}=139$~MeV at vacuum 
but also LQCD data~\cite{Karsch4,Karsch3,Kaczmarek2} on $\sigma$ and $|\Phi|$ 
for thermal systems with no $\mu_{\rm q}$ and $\mu_{\rm I}$~\cite{Sakai2}. 
The second set is $\Lambda =0.6315$~GeV, 
$G_{\rm s}=5.498$~GeV$^{-2}$, 
$G_{\rm s8}=0$ and $m_0=5.5$~MeV. 
This set can reproduce the pion decay constant 
and the pion mass  at vacuum, 
but not LQCD data for thermal systems 
with no $\mu_{\rm q}$ and $\mu_{\rm I}$. 
Thus, the first set with finite $G_{\rm s8}$ is more reliable.

\subsection{Average phase factor}
\label{Average phase factor}

As mentioned in section~\ref{Introduction}, we have to 
consider fluctuations to mean fields to 
evaluate the average phase factor. 
In this subsection, we consider static fluctuations (SF).

In the path-integral representation of the partition function $Z$ or 
the thermodynamic potential $\Omega$, 
$\phi_3$ and $\phi_8$ 
are fundamental fields rather than $\Phi$ and $\Phi^*$. 
This means that 
we should solve the stationary conditions   
\begin{eqnarray}
\frac{\partial\Omega}{\partial\varphi}=0
\label{stationary-condition}
\end{eqnarray}
for $\varphi =(\sigma, \vec{\pi}, \phi_3, \phi_8)$ rather than 
$\varphi =(\sigma, \vec{\pi}, \Phi, \Phi^*)$~\cite{Rossner}, 
although the solutions are not so different between the two cases. 
However, the first case does not guarantee that 
$\Omega$ is real. Following Ref.~\cite{Rossner}, 
we then put $\phi_8=0$ to keep $\Omega$ real. 
Noting that the first-order derivative with respect to $\phi_8$ 
does not vanish, 
we expand $\Omega$ up to quadratic terms of fluctuations: 
\begin{eqnarray}
\Omega=\Omega_{\rm MF}+\left(\frac{\delta\Omega}{\delta\varphi_i}\right)
_{\rm MF} \delta\varphi_i+
\frac{1}{2}\left(\frac{\delta^2\Omega}{\delta\varphi_i\delta\varphi_j}\right)
_{\rm MF} \delta\varphi_i\delta\varphi_j,~~~ 
\end{eqnarray}
where $\varphi=\sum_i \delta\varphi_i + \varphi_{\rm MF}$ 
for mean fields $\varphi_{\rm MF}$ and static (constant) 
fluctuations $\delta\varphi_i$. 
Since first-order terms in $\delta\varphi_i$ are purely imaginary, 
we can regard an integral over $\delta\varphi_i$ 
as a Fourier integral. We then obtain 
\begin{eqnarray}
Z&=&\int \prod_i d(\delta\varphi_i)\exp\left[-\frac{V}{T} \Omega \right]
=\frac{1}{\cal N} 
\exp\left[-\frac{V}{T}\tilde{\Omega} \right]
~~~~~~~
\label{eq:Z}
\end{eqnarray}
with 
\bea
\tilde{\Omega}=\left\{\Omega_{\rm MF}+\frac{1}{2}
\left(\frac{\delta^2\Omega}{\delta^2\phi_8}\right)^{-1}_{\rm MF}
\left(\frac{\delta\Omega}{\delta\phi_8}\right)^2_{\rm MF}\right\}
\eea
and  
\begin{eqnarray}
{\cal N}=\left(\frac{V}{2\pi T}\right)^{\frac{n}{2}}
\left\|\det H \right\|^{\frac{1}{2}}, ~~~~ 
\end{eqnarray}
where $n$ is the number of fields and $H$ is the 
Hessian matrix defined by
\begin{eqnarray}
H=\left[\frac{\delta^2\Omega}{\delta\varphi_i\delta\varphi_j}\right]_{\rm MF}. 
\label{Hessian}
\end{eqnarray}
Obviously, the Hessian matrix describes static fluctuations of 
mean fields $\varphi_{\rm MF}$. 
Therefore, the average phase factor is obtained by 
\eqref{MFgeneral} with $\Omega$ replaced by $\tilde{\Omega}$. 
The average phase factor thus obtained is a function of 
the $\varphi_{\rm MF}$ that satisfy the stationary conditions 
\eqref{stationary-condition}. 
Thus, the average phase factor is calculable 
in the MF+SF framework.

The static fluctuations are composed of those of $\sigma$ and 
$\pi$ and of $\phi_3$ and $\phi_8$. 
In subsection~\ref{meson-loop}, 
we keep treating static fluctuations for $\phi_3$ and $\phi_8$, 
but consider dynamical fluctuations for $\sigma$ and $\pi$, 
that is, $\sigma$ and $\pi$ mesons.

It was revealed in Ref.~\cite{Anderson} with the NJL model that 
in the pion condensate phase, where 
$\pi_{\rm MF}$ is finite and then a massless mode appears, 
the average phase factor vanishes owing to $\det H_{1+1^*}=0$. 
Obviously, this is true also for the PNJL model, 
as shown in \eqref{Hessian}. 
In the normal phase with no pion condensate, 
$\Omega_{1+1}$ and $\Omega_{1+1^*}$ are the same in the MF level, 
so that the average phase factor is determined by only the SF 
factor ${\sqrt{\det H_{1+1^*}}}/{\sqrt{\det H_{1+1}}}$.

LQCD calculation in Ref.~\cite{Elia2} has a lattice size $16^3\times 4$.  
Hence, the three-dimensional volume is $V=(16a)^3$ for a lattice spacing $a$ 
and the inverse of temperature is $1/T=4a$. 
The four-dimensional volume  is then 
obtained by $V/T=64T^{-4}$. This four-dimensional volume is taken 
also for the PNJL model in its calculation of the average phase factor.


\section{Numerical results}
\label{Numerical-results}

\subsection{Average phase factor and Polyakov loop}

Solving the stationary conditions \eqref{stationary-condition} and 
inserting the solutions in \eqref{MFgeneral}, 
one can obtain $\Phi$, $\pi$ and $\langle e^{2i\theta} \rangle$ 
as a function of $T$ and $\mu_{\rm q}$. 
As a shorthand notation, we use $\Phi$ for 
$\Phi_{1+1}$ and $\pi$ for $\pi_{1+1^*}$. 
All the calculations in this subsection 
are done without the scalar-type eight-quark interaction; 
roles of the eight-quark interaction will be discussed in subsection 
\ref{eight-quark}.

\begin{figure}[htbp]
\begin{center}
 \includegraphics[width=0.40\textwidth]{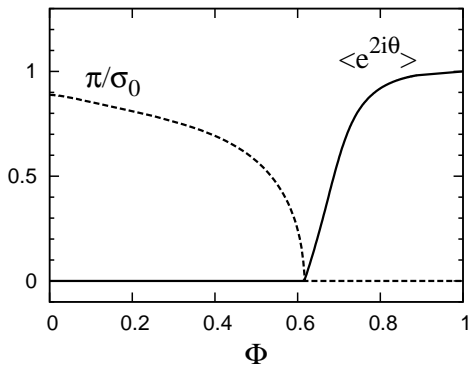}
\end{center}
\caption{$\Phi$ dependence of the scaled pion condensate 
$\pi/\sigma_0$ and the average phase factor 
$\langle e^{2i\theta}\rangle$ 
at $\mu_{\rm q}=100$~MeV. Here, $\sigma_0$ is a chiral condensate at 
$T=\mu_{\rm q}=0$. 
The former is plotted by a dashed curve, while the latter is by a solid 
curve. 
}
\label{fig1}
\end{figure}

In Fig.~\ref{fig1}, we plot $(\Phi (T),\pi (T))$ by a dashed line and 
$(\Phi (T),\langle e^{2i\theta} \rangle (T))$ by a solid line, 
varying $T$ with $\mu_{\rm q}$ fixed at 100~MeV. 
Since $\Phi$ is an increasing function of $T$~\cite{Sakai2}, 
an increase of $\Phi$ means that of $T$. 
The pion condensate $\pi$ decreases as $\Phi$ increases and 
finally vanishes at a critical value $\Phi_{\rm c}$ of $\Phi$. 
Below $\Phi_{\rm c}$, the average phase factor is always zero, 
while $\pi$ is finite. 
Above $\Phi_{\rm c}$, inversely, 
the average phase factor is finite, while $\pi$ is always zero. 
Thus, there is a negative correlation between the average phase 
factor and the pion condensate. This property 
is also seen in the NJL model~\cite{Anderson}. 
In contrast, there exists a positive correlation 
between the average phase factor and the Polyakov loop:   
the average phase factor is zero at small $\Phi$ such as 
$\Phi < \Phi_{\rm c}$, but at large $\Phi$ such as $\Phi > \Phi_{\rm c}$ 
the average phase factor is finite 
and an increasing function of $\Phi$. 
In the $\Phi=1$ limit, the average phase factor tends to 1. 
This implies that the NJL model overestimates the average phase factor 
compared with the PNJL model, 
since $\Phi=1$ in the NJL model. This is true, as shown below in 
Fig.~\ref{fig2}.

\begin{figure}[htbp]
\begin{center}
 \includegraphics[width=0.40\textwidth]{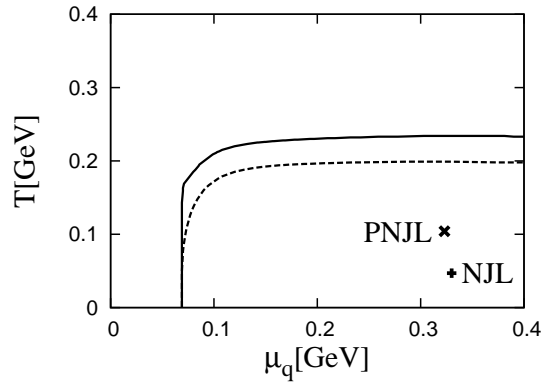}
\end{center}
\caption{ Boundary of the $\langle e^{2i\theta}\rangle=0$ region 
in $\mu_{\rm q}$-$T$ plane.  
A solid (dashed) line represents the boundary in the PNJL (NJL) model.  
Symbols $\times$ and $+$ stand for critical endpoints of 
the PNJL and NJL models, respectively. 
}
\label{fig2}
\end{figure}

Figure \ref{fig2} shows a boundary of the region where 
$\langle e^{2i\theta}\rangle =0$. The solid (dashed) curve is a result of 
the PNJL (NJL) model. 
The region is wider in the PNJL model compared with the NJL model. 
Thus, the fact that $\Phi < 1$ in the PNJL model makes the region 
wider and the average phase factor smaller outside the region. 
The critical endpoint (CEP) of the first-order chiral phase transition 
is also 
plotted by a cross ($\times$) for the PNJL model and a plus ($+$) for 
the NJL model. 
For both the models, thus, the CEP and the first-order phase transition are 
in the $\langle e^{2i\theta}\rangle =0$ region. 
This implies that the location of CEP cannot be determined by LQCD directly. 
The CEP is located at higher $T$ and lower $\mu_{\rm q}$ 
in the PNJL model compared with the NJL model. 
However, a relative distance of CEP to the boundary 
of the $\langle e^{2i\theta}\rangle =0$ region is almost the same between 
the two models.

\subsection{Scalar-type eight-quark inteaction}
\label{eight-quark}
The scalar-type eight-quark interaction
\begin{eqnarray}
G_{\rm s8}[(\bar{q}q)^2+(\bar{q}i\gamma_5\vec{\tau}q)^2]^2
\end{eqnarray}
is inevitable to reproduce LQCD data 
at finite real isospin chemical potential~\cite{Sasaki-T} 
and zero and finite imaginary quark chemical potential~\cite{Sakai2}. 
Furthermore, the interaction 
affects a location of CEP for real $\mu_{\rm q}$. 
Therefore, it is important to see how the scalar-type eight-quark interaction 
affects the average phase factor.

\begin{figure}[htbp]
\begin{center}
 \includegraphics[width=0.40\textwidth]{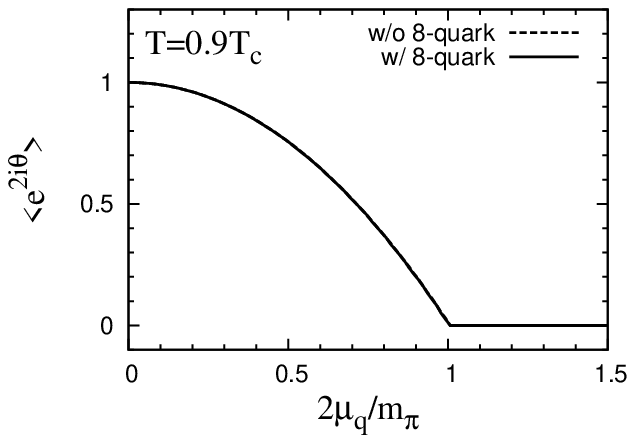}
 \includegraphics[width=0.40\textwidth]{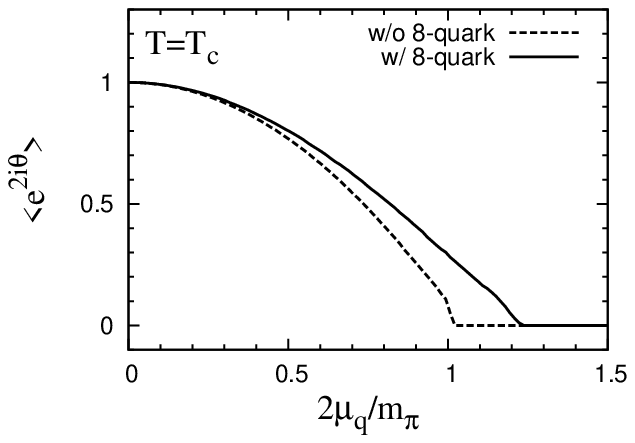}
 \includegraphics[width=0.40\textwidth]{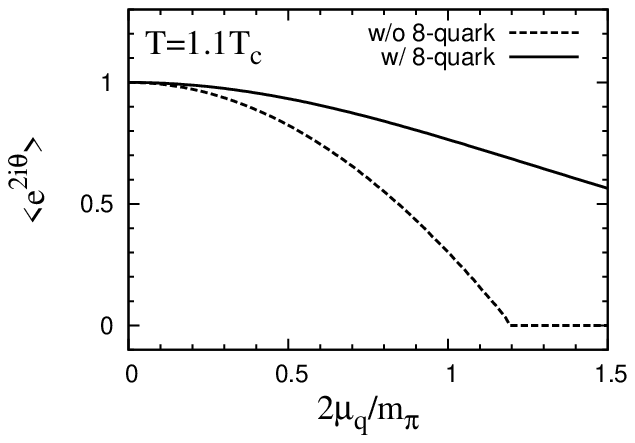}
\end{center}
\caption{Effect of scalar-type eight-quark interaction 
on average phase factor.
Solid and dashed lines stand for the PNJL result with and 
without the eight-quark interaction, respectively.
}
\label{Fig3}
\end{figure}

Figure~\ref{Fig3} shows $\mu_{\rm q}$ dependence of 
the average phase factor at $T=0.9T_{\rm c}, T_{\rm c}$ and $1.1T_{\rm c}$. 
The solid and dashed lines correspond to the PNJL result with and without 
the eight-quark interaction. 
Below $T_{\rm c}$ such as $T=0.9T_{\rm c}$, 
the scalar-type eight-quark interaction does not 
affect the average phase factor, and at $T=T_{\rm c}$ 
the effect becomes appreciable. Above $T_{\rm c}$ 
such as $T=1.1T_{\rm c}$, it enhances the average phase factor sizably.

\begin{figure}[htbp]
\begin{center}
 \includegraphics[width=0.40\textwidth]{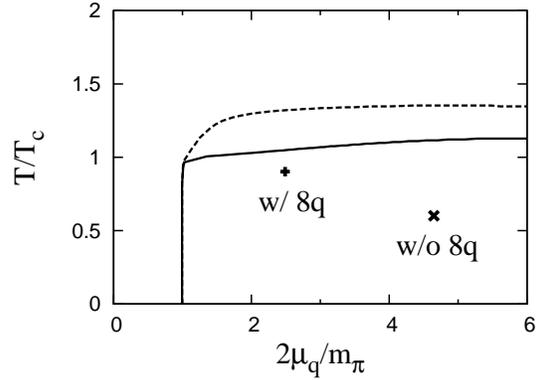}
\end{center}
\caption{ 
Effect of the scalar-type 8-quark interaction on 
locations of CEP and the boundary of 
the $\langle e^{2i\theta}\rangle=0$ region. 
Solid and dashed lines stand for the boundary calculated 
by the PNJL model with and 
without the eight-quark interaction, respectively.
Symbols $+$ and $\times$ mean the CEP calculated by 
the PNJL model with and without the eight-quark interaction, respectively. 
}
\label{fig4}
\end{figure}

Figure \ref{fig4} presents a boundary of 
the $\langle e^{2i\theta}\rangle=0$ region and 
locations of CEP calculated by the PNJL model 
without and with the scalar-type eight-quark interaction. 
The eight-quark interaction makes the region smaller. 
Meanwhile, it shifts the CEP to higher $T$ and lower $\mu_{\rm q}$. 
Thus, the relative distance of CEP to the boundary becomes much smaller 
by the scalar-type eight-quark interaction, 
although CEP itself lies in the $\langle e^{2i\theta}\rangle=0$ region even 
after the scalar-type eight-quark interaction is taken into account. 
If more accurate LQCD data becomes available in future outside 
the region $\langle e^{2i\theta}\rangle=0$, the PNJL model that reproduces 
the data can predict a location of CEP in principle. 
The reliability of the prediction may be 
proportional to the relative distance. 
In this sense, the fact that the relative distance is small is important.

\begin{figure}[htbp]
\begin{center}
 \includegraphics[width=0.40\textwidth]{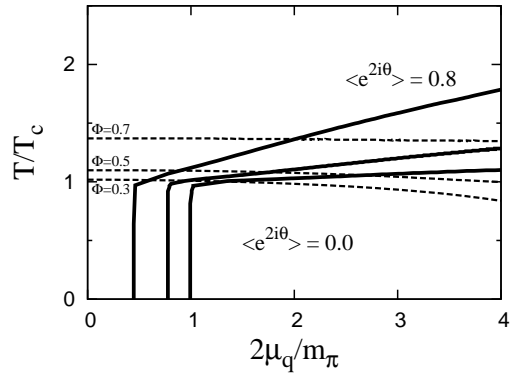}
\end{center}
\caption{Contours of the average phase factor and the Polyakov loop 
in $\mu_{\rm q}$-$T$ plane. 
Contours of $\langle e^{2i\theta}\rangle =0, 0.4, 0.8$ and 1 
are drawn by solid curves, while those of 
$\Phi=0.3, 0.5$ and $0.7$ are by dashed curves. 
}
\label{fig5}
\end{figure}

Figure~\ref{fig5} presents contours of $\langle e^{2i\theta}\rangle$ and 
$\Phi$ in $\mu_{\rm q}$-$T$ plane calculated by the PNJL model with 
scalar-type eight-quark interaction. 
Solid curves correspond to contours of 
$\langle e^{2i\theta}\rangle =0, 0.4,$ and 0.8, 
while dashed curves do to contours of 
$\Phi=0.3, 0.5$ and $0.7$. 
For $\mu_{\rm q} < m_{\pi}/2$, the average phase factor is 
finite and then calculable with LQCD in principle. 
For $\mu_{\rm q} > m_{\pi}/2$, $\langle e^{2i\theta}\rangle=0$ at 
$\Phi \la 0.5$ corresponding to the confinement region. 
At $\Phi \ga 0.5$ corresponding to the deconfinement region, 
the factor $\langle e^{2i\theta}\rangle$ is finite and 
an increasing function of $\Phi$, indicating that there is a 
positive correlation between $\langle e^{2i\theta}\rangle$ and $\Phi$.
Thus, LQCD is feasible only in the deconfinement phase, 
when $\mu_{\rm q} > m_{\pi}/2$.

\subsection{Dynamical mesonic fluctuations}
\label{meson-loop}

The average phase factor was calculated with LQCD~\cite{Elia2} 
in which the lattice size is 
$16^3\times 4$ and the pion mass at vacuum is $m_{\pi}^{\rm LQCD} \approx 280$~MeV. 
In the PNJL calculation, we have then varied the quark mass 
from $m_0=5.5$~MeV to $22.5$~MeV to reproduce $m_{\pi}^{\rm LQCD}=280$~MeV. 
For this value of $m_0$, the deconfinement transition temperature becomes 
a bit higher value, i.e., $T_{\rm c}=180$~MeV.

\begin{figure}[htbp]
\begin{center}
 \includegraphics[width=0.40\textwidth]{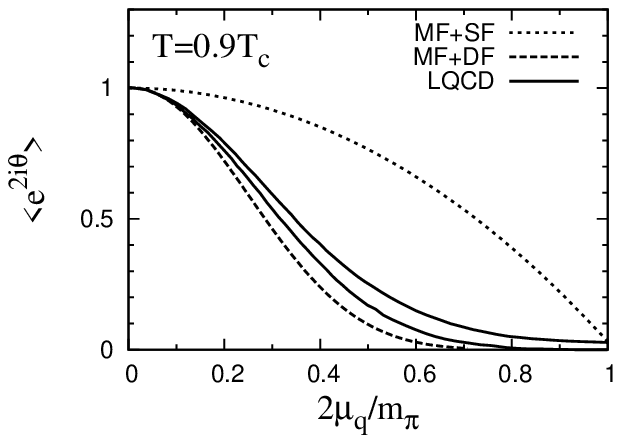}
 \includegraphics[width=0.40\textwidth]{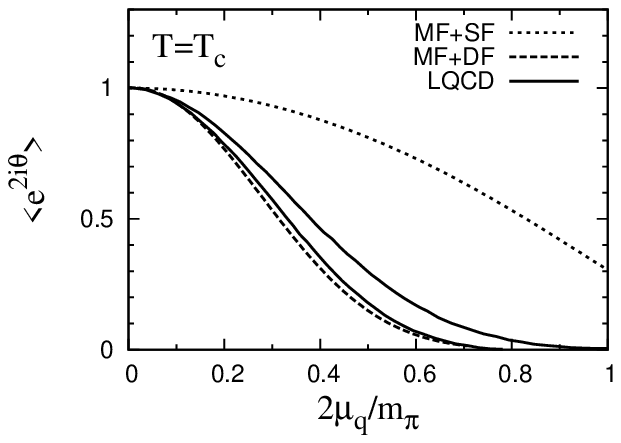}
 \includegraphics[width=0.40\textwidth]{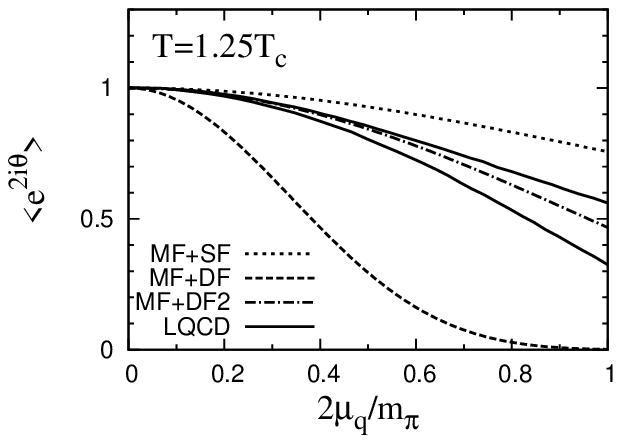}
\end{center}
\caption{Effects of mesonic fluctuations on the average phase factor. 
All the calculations take account of the eight-quark interaction. 
In MF+DF and MF+DF2 calculations, $m_\pi =m_{\pi}^{\rm LQCD}$ and 
$2.8m_{\pi}^{\rm LQCD}$, 
respectively.
}
\label{fig6}
\end{figure}

Figure \ref{fig6} shows $\mu_{\rm q}$ dependence 
of the average phase factor 
at $T=0.9T_{\rm c}, T_{\rm c}$ and $1.25T_{\rm c}$. 
LQCD result is evaluated from LQCD data at imaginary chemical by assuming 
a polynomial function. For each panel, the two solid lines 
delimit the 90\% confidence level region for the extrapolation, 
while the MF+SF calculation is represented by dotted lines. 
The MF+SF calculation overestimates LQCD data largely. 
Therefore, we consider dynamical mesonic fluctuations 
beyond the MF+SF framework.

Below $T_{\rm c}$ where the system is confined, 
it is natural to think that mesonic modes dominate rather than quark modes. 
Therefore, we should take all possible channels (mesonic modes) in 
the bubble summation in the random phase approximation (RPA). 
If there is no pion condensate, 
$\sigma$ and $\pi$ meson modes are decoupled to each other. 
Hence, the mesonic polarization function matrix 
does not have off-diagonal elements. 
Up to order $1/N_{\rm c}$, where $N_{\rm c}$ is the number of colors, 
the thermodynamic potential is obtained by 
\begin{eqnarray}
\Omega=\Omega_{\rm MF}+\Omega_{\rm DF}, 
\end{eqnarray}
where $\Omega_{\rm MF}$ is the mean-field part shown in \eqref{eq:E12-pi} 
and $\Omega_{\rm DF}$ is the dynamical mesonic fluctuation (DF) part 
in the ring diagram. 
Hence, $\Omega_{\rm DF}$ is obtained by~\cite{Zhuang} 
\begin{eqnarray}
\Omega_{\rm DF}=-\frac{i}{2}\int\frac{d^4q}{(2\pi)^4}
\ln\det[1-G^*_{\rm s}\Pi(q)] 
\label{Omegafl}
\end{eqnarray}
with the effective coupling 
$G_{\rm s}^*={\rm diag}(G^*_{{\rm s}j})$ for 
\begin{eqnarray}
G^*_{{\rm s}\sigma}=\frac{\partial M}{\partial \sigma},~~~
G^*_{{\rm s}\pi_+}=G^*_{{\rm s}\pi_-}=G^*_{{\rm s}\pi_0}=\frac{\partial N}{\partial \pi},
\end{eqnarray}
and the mesonic polarization bubbles 
\begin{eqnarray}
\Pi_{jk}(q)&=&i\int\frac{d^4p}{(2\pi)^4}
{\rm Tr}[\Gamma^*_j{\cal S}(p+q)\Gamma_k{\cal S}(p)]
\end{eqnarray}
for $j,~k=\sigma,~\pi_+,~\pi_-,~\pi_0$, 
where Tr is the trace in color, flavor and Dirac indexes and the four 
momentum integral is defined as 
$\int d^4q/(2\pi)^4=iT\sum_n\int d^3{\bf q}/(2\pi)^3$ for 
finite temperature. 
The meson vertexes $\Gamma_k$ depend on meson taken; precisely, 
$\Gamma_{\sigma}=1,~\Gamma_{\pi_+}=i\tau_+\gamma_5,
~\Gamma_{\pi_-}=i\tau_-\gamma_5,~\Gamma_{\pi_0}=i\tau_3\gamma_5$.

Before calculating dynamical mesonic fluctuations, we clarify 
a relation between static and dynamical mesonic fluctuations. 
The effective action of the NJL-type model to $1/N_{\rm c}$ order 
is derivable 
by using the auxiliary field method~\cite{Sakaguchi}. 
The second derivative 
of the effective action with respect to meson fields 
yields the inverse meson propagator
$G^*_{\rm s}[1-G^*_{\rm s}\Pi(q)]$. 
Since the static fluctuations are constant, 
the Hessian matrix (\ref{Hessian}) is obtained by 
setting the external momentum $q=0$ in 
the inverse propagator: 
\begin{eqnarray}
H=G^*_{\rm s}[1-G^*_{\rm s}\Pi(q=0)] . 
\end{eqnarray}
Setting $q=0$ in \eqref{Omegafl} reduces 
$\Omega_{\rm DF}$ to $\ln\det[H]$ and the resultant partition function 
turns out to be (\ref{eq:Z}). 
Thus, the static mesonic fluctuation $\Omega_{\rm SF} = \ln\det[H]$ 
can be regarded as an approximation to the dynamical one $\Omega_{\rm DF}$.

Now we introduce the effective meson mass $m_j^*$ which satisfy 
\begin{eqnarray}
\det[1-G_{\rm s}^*\Pi(q_0+\mu_j=m_j^*,{\bf q=0})]=0 
\end{eqnarray}
with the meson chemical potentials $\mu_{\sigma}=\mu_{\pi_0}=0$, 
$\mu_{\pi_+}=2\mu_{\rm I}$ and $\mu_{\pi_-}=-2\mu_{\rm I}$. 
Here note that physical meson masses are not $m_j^*$ 
but $m_j=m_j^*-\mu_j$ because they are calculated from $q_0$.

Since it is difficult to calculate the dynamical mesonic fluctuations 
(\ref{Omegafl}) exactly, 
here we make the pole approximation that neglects the scattering phase shift. 
If $T < T_{\rm c}$ and there is no pion condensation, 
$m_j^*$ is well approximated by the meson mass  $m_{j}^{0}$
at vacuum. In this approximation, $\Omega_{\rm DF}$ can be obtained 
by a sum of four quasiparticles, $\sigma$, $\pi^+$, $\pi^-$ and $\pi^0$: 
\begin{eqnarray}
\Omega_{\rm DF}&=&\sum_j\Omega_j,\\
\Omega_j&=&\int\frac{d^3{\bf q}}{(2\pi)^3}
\left[\frac{1}{2}(E_j-\mu_j)+T\ln\left(1-e^{-\beta(E_j-\mu_j)}\right)\right],
\notag \\
\end{eqnarray}
where $E_j=\sqrt{{\bf q}^2+{m^*}^2_j}$. 
However, corrections due to $\sigma$ and $\pi^0$ mesons are exactly 
cancelled out between $\Omega_{1+1}$ and $\Omega_{1+1^*}$. 
This framework is referred to as MF+DF in this paper; 
here, the static fluctuations of $\phi_3$ and $\phi_8$ 
are taken into account in the Hessian matrix.

As seen in Fig.~\ref{fig6}, 
at lower $T$ such as $T=0.9T_{\rm c}$ and $T_{\rm c}$, 
the MF+DF calculation (dashed line) almost reproduces LQCD data. 
Above $T_{\rm c}$ such as $T=1.25T_{\rm c}$,  
the MF+DF calculation underestimates LQCD data. 
For $T>T_{\rm c}$, in general, the pion mass becomes larger than  
$m^0_{\pi}$~\cite{Xiong}. 
If $m_\pi = 2.8m^0_{\pi}$ is taken, the calculation (dot-dashed line) 
is consistent with LQCD data; this calculation is denoted by MF+DF2 
in Fig.~\ref{fig6}.

\section{Summary}
\label{Summary}

We have calculated the average phase factor of the QCD determinant 
at finite quark chemical potential $\mu_{\rm q}$, 
using the two-flavor version of 
the PNJL model 
with the scalar-type eight quark interaction, 
since the model is successful in reproducing 
LQCD data 
not only on the 1+1 system in the limit of no $\mu_{\rm q}$~\cite{Sakai2}
but also on the $1$+$1^*$ system with finite isospin chemical 
potential $\mu_{\rm I}$~\cite{Sasaki-T}. 
In the present model, there exists a critical endpoint (CEP) in the 
1+1 system with finite $\mu_{\rm q}$. The CEP lies 
inside the $\langle e^{2i\theta}\rangle=0$ region. This implies 
that the location cannot be determined by LQCD solely.

For $\mu_{\rm q} > m_{\pi}/2$, the pion condensate occurs at 
$\Phi \la 0.5$, so that $\langle e^{2i\theta}\rangle=0$ there. 
At $\Phi \ga 0.5$, the factor $\langle e^{2i\theta}\rangle$ is finite and 
an increasing function of $\Phi$. Thus, there 
exists a positive correlation 
between $\langle e^{2i\theta}\rangle$ and $\Phi$. 
Therefore, LQCD is feasible 
only in the deconfinement phase with large $\Phi$, 
when $\mu_{\rm q} > m_{\pi}/2$.

The eight-quark interaction makes the $\langle e^{2i\theta}\rangle=0$ region 
shrink, while 
it shifts the CEP toward higher $T$ and smaller $\mu_{\rm q}$. 
As a consequence of these properties, a relative distance of the CEP to 
the boundary of the $\langle e^{2i\theta}\rangle=0$ region becomes smaller. 
If more accurate LQCD data becomes available in future outside 
the $\langle e^{2i\theta}\rangle=0$ region, 
we can predict a location of the CEP with 
the PNJL model the parameters of which are fitted to the data. 
The accuracy of the model prediction seems to be 
proportional to the distance. 
In this sense, the fact that the relative distance is small is important.

For $\mu_{\rm q} < m_{\pi}/2$ where no pion condensate takes place, 
the PNJL calculation with static fluctuations 
cannot reproduce LQCD data~\cite{Elia2} 
at both $T \le T_{\rm c}$ and $T>T_{\rm c}$. 
This problem is solved partly by treating dynamical mesonic fluctuations with 
the pole approximation. 
The PNJL model with the dynamical mesonic fluctuations 
reproduces LQCD data at $T \le T_{\rm c}$. 
For $T>T_{\rm c}$, however, the calculation cannot reproduce the data. 
The first possible reason is 
that the meson mass at $T>T_{\rm c}$ is different from the value at vacuum. 
The second possible reason is that the pole approximation is not good, 
because meson at $T>T_{\rm c}$ is generally in a resonance state. 
The third possible reason is effects of physical states 
with nonzero baryon numbers.  
It is reported in Ref.~\cite{Elia2} that the physical states 
tend to enhance the average phase factor, although 
the PNJL model does not include such effects. 
This problem should be solved in future in order to construct 
a reliable effective model that makes it possible to predict a location of 
CEP and the phase diagram at finite $\mu_{\rm q}$.

\noindent
\begin{acknowledgments}
The authors thank K. Kashiwa for useful discussions and suggestions. 
H. K. also thanks M. Imachi, H. Yoneyama, H. Aoki and M. Tachibana for usefull 
discussions. 
Y. S. is supported by JSPS Research Fellow.
\end{acknowledgments}


\end{document}